\documentclass[a4paper,12pt]{article}
\usepackage{amsmath,amssymb,graphicx,appendix,color}
\RequirePackage[sort&compress,square,comma,numbers]{natbib}

\allowdisplaybreaks[1]
\newcommand{\T}{\textbf{\textit{T}}}
\newcommand{\Z}{\textbf{\textit{Z}}}
\newcommand{\ep}{\epsilon}
\newcommand{\ord}{{\cal{O}}}
\newcommand{\as}{\alpha_s}

\begin{document}

\begin{titlepage}

\begin{flushright}
\normalsize
MZ-TH/12-36\\
August 23, 2012
\end{flushright}

\vspace{0.3cm}
\begin{center}
\Large\bf
Structure of Infrared Singularities of Gauge-Theory Amplitudes at Three and Four Loops
\end{center}

\vspace{0.8cm}
\begin{center}
{\sc Valentin Ahrens, Matthias Neubert and Leonardo Vernazza}\\
\vspace{0.7cm}
{\sl PRISMA Cluster of Excellence \& Institut f\"ur Physik (THEP)\\ 
Johannes Gutenberg-Universit\"at, D--55099 Mainz, Germany}
\end{center}

\vspace{0.8cm}
\begin{abstract}
The infrared divergences of massless $n$-parton scattering amplitudes can be derived from the anomalous dimension of $n$-jet operators in soft-collinear effective theory. Up to three-loop order, the latter has been shown to have a very simple structure: it contains pairwise color-dipole interactions among the external partons, governed by the cusp anomalous dimension and a logarithm of the kinematic invariants $s_{ij}$, plus a possible three-loop correlation involving four particles, which is described by a yet unknown function of conformal cross ratios of kinematic invariants. This function is constrained by two-particle collinear limits and by the known behavior of amplitudes in the high-energy limit. We construct a class of relatively simple functions satisfying these constraints. We also extend the analysis to four-loop order, finding that three additional four-particle correlations and a single five-particle correlation appear, which again are governed by functions of conformal cross ratios. Our results suggest that the dipole conjecture, which states that only two-particle color-dipole correlations appear in the anomalous dimension, may need to be generalized. We present a weaker form of the conjecture, stating that to all orders in perturbation theory corrections to the dipole formula are governed by functions of conformal cross ratios, and are ${\cal O}(1/N_c^2)$ suppressed relative to the dipole term. If true, this conjecture implies that the cusp anomalous dimension obeys Casimir scaling to all orders in perturbation theory.
\end{abstract}
\vfil

\end{titlepage}
\newpage

\section{Introduction}

Understanding the structure of infrared (IR) singularities of gauge-theory scattering amplitudes is an important problem. On one hand, it helps in unveiling the deeper structure of quantum field theory in higher orders of perturbation theory; on the other, it also has practical applications. The ability to predict the IR singularities of $n$-parton amplitudes enables one to systematically resum large logarithmic corrections to cross sections and differential distributions for many important collider processes. This leads to a higher precision in the calculation of these observables. 

The problem of predicting the structure of IR singularities of on-shell $n$-particle scattering amplitudes in massless QCD simplifies, if one realizes that they can be put in one-to-one correspondence with ultraviolet (UV) divergences of operators defined in soft-collinear effective theory (SCET) \cite{Becher:2009cu}. This idea implies that IR divergences can be studied by means of standard renormalization-group techniques -- a concept that had been developed earlier in the context of theories of Wilson lines \cite{Korchemskaya:1994qp}. The IR divergences of $n$-point scattering amplitudes can be absorbed into a multiplicative renormalization factor $\Z$, which can be derived from an anomalous dimension $\bf\Gamma$. Both objects are matrices in color space, i.e.\ they mix amplitudes with the same particle content but different color structures. The predictive power of this approach relies on the fact that the anomalous dimension $\bf\Gamma$ is tightly constrained by the structure of the effective field theory \cite{Becher:2009cu}: soft-collinear factorization implies that $\bf\Gamma$ splits into a collinear and a soft contribution, ${\bf\Gamma}={\bf\Gamma}_c+{\bf\Gamma}_s$, and given that there are no interactions among different collinear sectors of SCET, all non-trivial color and momentum dependence is encoded in the soft anomalous dimension ${\bf\Gamma}_s$. 

The quantity ${\bf\Gamma}_s$ is the anomalous dimension associated with an operator built out of $n$ Wilson lines, one for each external parton, directed along the momentum of that parton and living in the appropriate representation of $SU(N_c)$. We use the color-space formalism, in which amplitudes are treated as $n$-dimensional vectors in color space \cite{Catani:1996vz}. $\T_i$ is the color generator associated with the $i$-th parton in the scattering amplitude, which acts as an $SU(N_c)$ matrix on the color indices of that parton. Explicitly, one has $(\T_i^a)_{\alpha\beta}= t^a_{\alpha\beta}$ for a final-state quark or initial-state anti-quark, $(\T_i^a)_{\alpha\beta}=-t^a_{\beta\alpha}$ for a final-state anti-quark or initial-state quark, and $(\T_i^a)_{bc}=-if^{abc}$ for a gluon. The dependence of the soft anomalous dimension on the external momenta $p_i$ of the partons is encoded via so-called cusp angles $\beta_{ij}$ (with $i\ne j$), which for slightly off-shell, massless partons are defined as
\begin{equation}
   \beta_{ij} = \ln\frac{(-s_{ij})\,\mu^2}{(-p_i^2) (-p_j^2)} \,,
\end{equation}
where $s_{ij}=2\sigma_{ij}\,p_i\cdot p_j+i0$, and the sign factor $\sigma_{ij}=+1$ if the momenta $p_i$ and $p_j$ are both incoming or outgoing, and $\sigma_{ij}=-1$ otherwise. The dependence on the collinear regulators $p_i^2$ disappears in the sum ${\bf\Gamma}={\bf\Gamma}_c+{\bf\Gamma}_s$, such that the complete anomalous dimension $\bf\Gamma$ only depends on the kinematic invariants $s_{ij}$. Here \cite{Becher:2003kh}
\begin{equation}\label{Gammac}
   {\bf\Gamma}_c = \sum_i\,\Big[ - \Gamma^i_{\rm cusp}(\as)\,\ln\frac{\mu^2}{-p_i^2}
    + \gamma_c^i(\as) \Big] 
\end{equation}
contains the sum of the collinear contributions, where $\Gamma^i_{\rm cusp}(\as)$ is the cusp anomalous dimension in the color representation of parton $i$ \cite{Korchemskaya:1992je}. 

The structure of the soft anomalous dimension is constrained in three different ways:
\begin{itemize}
\item 
Soft-collinear factorization, the fact that the interactions between soft and collinear particles in the SCET Lagrangian can be removed by a field redefinition \cite{Bauer:2001yt}, implies a set of partial differential equations \cite{Gardi:2009qi,Becher:2009qa}, which can be written in the form
\begin{equation}\label{eq9}
   \sum_{j\ne i}\,\frac{\partial{\bf\Gamma}_s(\{\underline \beta\},\mu)}{\partial \beta_{ij}} 
   = \Gamma^i_{\rm cusp}(\as) \,,
\end{equation}
where $\{\underline{\beta}\}=\{\beta_{12},\dots,\beta_{ij},\dots\beta_{n-1,n}\}$ denotes the set of cusp angles. These equations allow two types of dependences on the cusp angles: either a linear dependence on $\beta_{ij}$, or an arbitrary dependence on the conformal cross ratios
\begin{equation}\label{betadef}
   \beta_{ijkl} \equiv \beta_{ij} + \beta_{kl} - \beta_{ik} - \beta_{jl} 
   = \ln\frac{(-s_{ij}) (-s_{kl})}{(-s_{ik}) (-s_{jl})} 
   \equiv \ln\rho_{ijkl} \,.
\end{equation}   
The latter possibility is allowed because the differential operator in (\ref{eq9}) gives zero when acting on a conformal cross ratio, so any function of conformal cross ratios is a solution to the homogeneous equation associated with (\ref{eq9}).
\item 
The non-abelian exponentiation theorem \cite{Gatheral:1983cz,Frenkel:1984pz} implies that only single connected gluon webs, whose ends can be attached in arbitrary ways to the $n$ Wilson lines, contribute to the soft anomalous dimension \cite{Gardi:2009qi,Becher:2009qa}. This imposes tight constraints on the color structures that can arise in higher orders of the loop expansion. The generalization of the concept of ``webs'' to multi-parton amplitudes has been discussed in detail in \cite{Gardi:2010rn}.
\item 
In the limit where two or more partons become collinear, an $n$-parton scattering amplitude splits into an ($n-1$)-parton amplitude times a process-independent splitting amplitude, which involves the collinear partons only \cite{Berends:1988zn,Mangano:1990by,Bern:1995ix,Kosower:1999xi}. The fact that the anomalous dimension of the splitting amplitude must be independent of the momenta and color generators of the partons not involved in the splitting process imposes a non-trivial constraint on ${\bf\Gamma}_s$ \cite{Becher:2009qa}. 
\end{itemize}
Up to three-loop order, the most general form of the soft anomalous dimension compatible with these constraints reads \cite{Becher:2009qa}
\begin{equation}\label{eq8}
\begin{aligned}
   {\bf\Gamma}_s(\{\underline \beta\},\mu) 
   &= - \sum_{(i,j)}\,\frac{\T_i\cdot\T_j}{2}\,\gamma_{\rm cusp}(\as)\,\beta_{ij}
    + \sum_i \gamma_s^i(\as) \\
   &\quad\mbox{}+ \sum_{(i,j,k,l)} {\cal T}_{ijkl}\,F(\beta_{ijkl},\beta_{iklj}-\beta_{iljk})
    + \ord(\as^4) \,,
\end{aligned}
\end{equation}
where we use the short-hand notations $\T_i\cdot\T_j=\T_i^a\T_j^a$ (summed over $a$) and ${\cal T}_{ijkl}=f^{ade} f^{bce} (\T_i^a\T_j^b\T_k^c\T_l^d)_+$.\footnote{We define the symmetrized product $(\T_{i_1}^{a_1}\dots\T_{i_n}^{a_n})_+ =\frac{1}{n!} \sum_{\sigma\in S_n} \T_{i_{\sigma(1)}}^{a_{\sigma(1)}}\dots\T_{i_{\sigma(n)}}^{a_{\sigma(n)}}$, where $S_n$ is the set of permutation of $n$ objects.}
Under index permutations, the structure ${\cal T}_{ijkl}$ behaves in exactly the same way as the conformal cross ratio $\beta_{ijkl}$ in (\ref{betadef}). In the equation above the sums run over the $n$ external partons. The notation $(i_1,\dots,i_k)$ refers to unordered tuples of distinct parton indices. The coefficient functions $\gamma_{\rm cusp}(\as)$ and $\gamma_s^i(\as)$ have a perturbative series in $\as$ starting at one-loop order. The latter quantity depends on whether the parton $i$ is a quark or a gluon. The universal function $\gamma_{\rm cusp}(\as)$ is related to the cusp anomalous dimension of parton $i$ by the Casimir-scaling relation
\begin{equation}\label{casimir}
   \Gamma_{\rm cusp}^i(\as) = C_i\,\gamma_{\rm cusp}(\as) + \ord(\as^4) \,,
\end{equation}
where $C_i=\T_i^2$ equals $C_F$ for a particle in the fundamental representation ($i=q,\bar q$) and $C_A$ for one in the adjoint representation ($i=g$) of the gauge group. It is expected on general grounds that Casimir scaling is not an exact property of the cusp anomalous dimension. Indeed, using arguments based on the AdS/CFT correspondence, results for the cusp anomalous dimension obtained in the strong-coupling limit were found to be inconsistent with relation (\ref{casimir}) \cite{Armoni:2006ux,Alday:2007hr,Alday:2007mf}. In perturbation theory, Casimir scaling could first be violated at four-loop order. We will come back to this question below. The function $F(x,y)$ in the last term in (\ref{eq8}) has a perturbative series starting at three-loop order (or later). Its dependence on $\as$ is suppressed in our notation for simplicity.

We finally also quote the result for the full anomalous dimension ${\bf\Gamma}={\bf\Gamma}_c+{\bf\Gamma}_s$, which is obtained from (\ref{eq8}) by adding the collinear contribution (\ref{Gammac}). One obtains
\begin{equation}\label{eq8a}
\begin{aligned}
   {\bf\Gamma}(\{\underline p\},\mu) 
   &= \sum_{(i,j)}\,\frac{\T_i\cdot\T_j}{2}\,\gamma_{\rm cusp}(\as)\,\ln\frac{\mu^2}{-s_{ij}}
    + \sum_i \gamma^i(\as) \\
   &\quad\mbox{}+ \sum_{(i,j,k,l)} {\cal T}_{ijkl}\,F(\beta_{ijkl},\beta_{iklj}-\beta_{iljk})
    + \ord(\as^4) \,,
\end{aligned}
\end{equation}
which very closely resembles the expression for the soft anomalous dimension. Here $\gamma^i=\gamma_s^i+\gamma_c^i$. The right-hand side now only depends on the kinematic invariants $s_{ij}$. In Appendix~\ref{ap.1} we show explicitly how this anomalous dimension determines the IR poles of $n$-parton scattering amplitudes up to four-loop order.

The terms shown in the first line of (\ref{eq8}) and (\ref{eq8a}) involve only pairwise correlations among the color charges and momenta of the different partons. These are the familiar color-dipole correlations arising already at one-loop order from a single gluon exchange. These terms provide a solution to the inhomogeneous partial differential equations (\ref{eq9}), if one assumes that the Casimir-scaling relation (\ref{casimir}) holds to all orders in perturbation theory. The {\em dipole conjecture\/} \cite{Becher:2009cu,Gardi:2009qi,Becher:2009qa} states that this is indeed the case, and that to all orders the anomalous dimension might indeed be given by just the two terms shown in the first line of (\ref{eq8}) and (\ref{eq8a}). If this conjecture holds true, then this would indicate a semi-classical origin of IR singularities. Starting at three-loop order, however, the four-parton correlation term given in the second line of (\ref{eq8}) and (\ref{eq8a}) is allowed by the constraints summarized above, provided that the function $F$ vanishes in all collinear limits \cite{Becher:2009qa,Gardi:2009qi,Dixon:2009ur}. Note that given four different parton indices $i,j,k,l$, there are only two linearly independent conformal cross ratios, since 
\begin{equation}\label{lindep}
   \beta_{ijkl} + \beta_{iklj} + \beta_{iljk} = 0 \,,
\end{equation}
and all other cross ratios are related to the ones above by means of the symmetry relations $\beta_{ijkl}=\beta_{jilk}=-\beta_{ikjl}=-\beta_{ljki}=\beta_{klij}$. Following \cite{Becher:2009qa}, we choose the arguments of the function $F$ such that they match the symmetry properties of the color structure ${\cal T}_{ijkl}$. It then follows that the function $F(x,y)$ must be odd in its first argument.

It has recently been shown that the high-energy (``Regge'') limit imposes an interesting additional constraint on $n$-parton scattering amplitudes, which has important implications for the functional dependence of the anomalous dimension $\bf\Gamma$ on the conformal cross ratios $\beta_{ijkl}$ \cite{Bret:2011xm,DelDuca:2011ae}. The point is that the leading IR singularities of the Regge slopes are correctly described by the dipole conjecture, so extra contributions from functions such as $F$ in (\ref{eq8a}) must only give rise to subleading logarithms. Interestingly, this condition is not fulfilled for any of the candidate functions for $F$ proposed in the literature \cite{Dixon:2009ur}. On the other hand, it has not been demonstrated that the Regge constraint is strong enough to exclude the existence of non-zero functions of conformal cross ratios.

The goal of this paper is to push the analysis of the anomalous dimension further by addressing a couple of open issues. In Section~\ref{a.2}, we answer the question whether the Regge constraint excludes the existence of multi-parton correlations in the anomalous dimension by constructing explicit examples for the function $F$ in (\ref{eq8}) and (\ref{eq8a}), which vanish in all collinear limits and in addition do not give rise to leading logarithms in the high-energy limit. This proves that such functions can exist, and that therefore the Regge constraint does not help to simplify the structure of the anomalous dimension beyond the constraints imposed by soft-collinear factorization, non-abelian exponentiation, and collinear limits. In Section~\ref{a.3}, we extend the analysis of the anomalous dimension $\bf\Gamma$ presented in \cite{Becher:2009qa} to the level of four loops, thereby gaining further insights into the structure of the result in higher orders. At $\ord{(\as^4)}$ new webs involving four and five gluons appear and give rise to interesting new color structures. We confirm the finding of \cite{Becher:2009qa}, that even at four-loop order the contributions to the soft anomalous dimension ${\bf\Gamma}_s$ that are linear in the cusp angles still have the structure shown in the first line in (\ref{eq8}). In addition, we find four new functions of conformal cross ratios, accompanied by color structures correlating four or five external partons. As a by-product of our analysis, we extend the analysis of two-parton collinear limits performed in \cite{Becher:2009qa} to the more general case where three or more parton momenta become collinear. As discussed in Appendix~\ref{ap.2}, these multi-parton collinear limits do not provide additional constraints at three-loop order, but they may yield useful information on one of the new functions of conformal ratios associated with five-gluon webs, which appear first at the level of four loops. Our conclusions are presented in Section~\ref{a.4}.

\section{Consistent examples of four-parton correlations}
\label{a.2}

It has recently been shown that the ``Reggeization'' of scattering amplitudes in the high-energy limit can be used to derive a non-trivial constraint on the functional dependence of the four-parton correlation term $F$ in (\ref{eq8}) and (\ref{eq8a}) on the conformal cross ratios $\beta_{ijkl}$ \cite{Bret:2011xm,DelDuca:2011ae}. In the limit in which the center-of-mass energy $\sqrt{s}$ is much larger than the momentum transfer $\sqrt{-t}$ in the process, i.e.\ $|s/t|\to\infty$ at fixed $t$, amplitudes for $2\to n$ scattering processes are dominated by $t$-channel exchanges of particles, whose propagators get dressed according to the generic form
\begin{equation}
   \frac{1}{t} \to \frac{1}{t} \left( \frac{s}{-t} \right)^{\alpha_i(t)} , 
\end{equation}
where $\alpha_i(t)$ is referred to as the Regge trajectory of particle $i$. In this process, large logarithms $L\equiv\ln|s/t|\gg 1$ are resummed to leading logarithmic order, as has been proved for the cases of the exchange of a gluon \cite{Balitsky:1979ap} and a quark \cite{Bogdan:2006af}. For the gluon case, Reggeization of the cross section has also been proved at next-to-leading logarithmic order \cite{Fadin:2006bj}. The key observation made in \cite{Bret:2011xm,DelDuca:2011ae}, extending earlier work using Wilson lines \cite{Korchemskaya:1996je}, was that Regge trajectories are IR divergent in perturbation theory, and it is therefore possible to study them, and the high-energy limit in general, using our understanding of the structure of IR divergences as captured by (\ref{eq8a}). It was observed that the dipole formula -- the terms shown in the first line of (\ref{eq8a}) -- correctly reproduces the known behavior of the Regge trajectories of gluons and quarks at leading logarithmic order, and in fact it extends it to particles transforming under an arbitrary representation of the gauge group. Moreover, it was shown that at next-to-leading logarithmic order Reggeization still holds for the real part of the scattering amplitude, but that it fails for the imaginary part. At next-to-next-to-leading logarithmic order, the breakdown of Reggeization is a generic feature of scattering amplitudes. 

These results imply interesting constraints on the anomalous dimension $\bf\Gamma$ \cite{Bret:2011xm,DelDuca:2011ae}. To see this, consider for simplicity a $2\to 2$ scattering process in the high-energy limit $s\gg|t|$, where $s=s_{12}=s_{34}$ and $t=s_{13}=s_{24}$. Using color conservation, relation (\ref{eq8a}) yields for this case
\begin{equation}
\begin{aligned}
   {\bf\Gamma}(\{\underline p\},\mu) 
   &= \gamma_{\rm cusp}(\as) \left[ \T_t^2\,\ln\frac{s+t}{-t}
    + \T_s^2 \left( i\pi + \ln\frac{s+t}{s} \right) \right] \\
   &\quad\mbox{}+ \sum_{i=1}^4 \left[ \gamma^i(\as) - \frac{C_i}{2}\,\gamma_{\rm cusp}(\as) 
    \left( \ln\frac{\mu^2}{-t} + i\pi + \ln\frac{s+t}{s} \right) \right] \\
   &\quad\mbox{}+ 8 {\cal T}_{1234}\,\big[
    F(\beta_{1234},\beta_{1342}-\beta_{1423}) - F(\beta_{1423},\beta_{1234}-\beta_{1342}) \big] \\
   &\quad\mbox{}+ 8 {\cal T}_{1342} \big[ 
    F(\beta_{1342},\beta_{1423}-\beta_{1234}) - F(\beta_{1423},\beta_{1234}-\beta_{1342}) \big] \,,
\end{aligned}
\end{equation}
where we have introduced the $s$- and $t$-channel color operators $\T_s=\T_1+\T_2$ and $\T_t=\T_1+\T_3$ \cite{Dokshitzer:2005ig}. The values of the relevant conformal cross ratios are
\begin{equation}
   \beta_{1234} = 2 \left( \ln\frac{s}{-t} - i\pi \right) , \qquad
   \beta_{1342} = -2 \ln\frac{s+t}{-t} \,, \qquad
   \beta_{1423} = 2 \left( \ln\frac{s+t}{s} + i\pi \right) .
\end{equation}
In the high-energy limit, neglecting exponentially small terms in $L=\ln|s/t|$, we find from above
\begin{equation}\label{2to2}
\begin{aligned}
   {\bf\Gamma}(\{\underline p\},\mu) 
   &\to \left( L\T_t^2 + i\pi\T_s^2 \right) \gamma_{\rm cusp}(\as) 
    + \sum_{i=1}^4 \left[ \gamma^i(\as) - \frac{C_i}{2}\,\gamma_{\rm cusp}(\as)
    \left( \ln\frac{\mu^2}{-t} + i\pi \right) \right] \\
   &\quad\mbox{}+ 8{\cal T}_{1234}\,\big[ F(2L-2i\pi,-2L-2i\pi) - F(2i\pi,4L-2i\pi) \big] \\
   &\quad\mbox{}+ 8{\cal T}_{1342} \big[ F(-2L,-2L+4i\pi) - F(2i\pi,4L-2i\pi) \big] \,.
\end{aligned}
\end{equation}
The first term in this result, proportional to $L\T_t^2\,\gamma_{\rm cusp}(\as)$, correctly reproduces the leading logarithms in the Regge trajectories of any particle exchanged in the $t$-channel. The second term, $i\pi\T_s^2\,\gamma_{\rm cusp}(\as)$, is a non-universal source of breaking of Reggeization at next-to-leading logarithmic order, since in general the color structure $\T_s^2$ is not diagonal in the $t$-channel color basis. However, this breaking only contributes to the imaginary part of the amplitude. Likewise, we conclude that the terms in the last two lines, which multiply independent color structures times loop functions that arise first at three-loop order,  must not contain terms proportional to $\as^3 L^3$ or higher in the limit $L\to\infty$. Terms of $\ord{(i\as^3 L^2)}$ are allowed, since they break Reggeization only in the imaginary part of the amplitude, as are generic terms of $\ord{(\as^3 L)}$. Interestingly, as has already been pointed out in \cite{Bret:2011xm}, this condition is not satisfied for any of the example functions constructed in \cite{Dixon:2009ur}, all of which contain super-leading $L^4$ terms. 

It is, however, not too difficult to construct functions similar to those proposed in \cite{Dixon:2009ur}, which vanish in the two-particle collinear limit and contain only subleading logarithms in the Regge limit. Since the relevant three-loop web is the same as in ${\cal N}=4$ super Yang-Mills theory, these functions should have maximum transcendentality $\tau=5$. A simple class of such functions contains products of logarithms and dilogarithms of the variables $\rho_{ijkl}=e^{\beta_{ijkl}}$ defined in (\ref{betadef}). Specifically, consider the functions
\begin{equation}\label{ansatz}
   f_{n_1 n_2}(\rho_{ijkl},\rho_{iklj},\rho_{iljk}) 
   = \frac{\ln\rho_{ijkl}}{2n_1^2 n_2^2}\,\big[ g(\rho_{iklj}^{n_1})\,g(\rho_{iljk}^{n_2})
    + g(\rho_{iklj}^{n_2})\,g(\rho_{iljk}^{n_1}) \big] \,,
\end{equation}
where
\begin{equation}
\begin{aligned}
   g(z) &= \mbox{Li}_2(1-z) - \mbox{Li}_2(1-z^{-1}) \\
   &= \frac12\ln^2 z + \frac{\pi^2}{3} - 2\ln z \ln(1-z) - 2\mbox{Li}_2(z) \,.
\end{aligned}
\end{equation}
The definition in the first line makes explicit that this function is odd under the reflection $z\leftrightarrow 1/z$, while the form shown in the second line is convenient for deriving the asymptotic behavior for small $z$. The functions $f_{n_1 n_2}$ are chosen such that they vanish in the relevant collinear limits. As shown in \cite{Becher:2009qa}, these are (i) $\rho_{12kl}\to 0$, $\rho_{1kl2}\to 1$, $\rho_{1l2k}\to 1/\rho_{12kl}\to\infty$, and (ii) $\rho_{1jk2}\to 1$, $\rho_{1k2j}\to 1/\rho_{12jk}\to\infty$, $\rho_{12jk}\to 0$. For the $\rho$ variables approaching 1 in the collinear limit the corresponding functions $\ln\rho$ or $g(\rho^n)$ vanish like a power of $p_\perp\to 0$, while the remaining functions diverge like a power of $\ln p_\perp$. Expressing the $\rho_{ijkl}$ variables in terms of the arguments of the function $F(x,y)$ in (\ref{eq8a}), we are led to consider the class of functions
\begin{equation}
   F_{n_1 n_2}(x,y) = f_{n_1 n_2}(e^x,e^{-\frac12(x-y)},e^{-\frac12(x+y)}) \,,
\end{equation}
which as required are odd in their first argument. We find that with this ansatz only one of the three functions entering in (\ref{2to2}) is non-zero in the Regge limit:
\begin{equation}
\begin{aligned}
   &F_{n_1 n_2}(2L-2i\pi,-2L-2i\pi) \to 0 \,, \\
   &F_{n_1 n_2}(-2L,-2L+4i\pi) \to 0 \,, \\
   &F_{n_1 n_2}(2i\pi,4L-2i\pi) \\
   &\to -4i\pi \left[ 2L^2(L-i\pi)^2 + \left( \frac{1}{n_1^2} + \frac{1}{n_2^2} \right)
    \left( L^2 + (L-i\pi)^2 \right) + \frac{\pi^4}{18 n_1^2 n_2^2} \right] .
\end{aligned}
\end{equation}
It follows that any difference of two such functions is free of the super-leading $L^4$ and leading $L^3$ terms, and starts with a subleading logarithms $iL^2$, which is allowed by all known properties of Regge factorization. Hence, any function
\begin{equation}\label{testfun}
   F(x,y) = \sum_{n_1,n_2}\,a_{n_1 n_2}\,F_{n_1 n_2}(x,y) \,, \qquad
   \mbox{with} \quad
   \sum_{n_1,n_2}\,a_{n_1 n_2} = 0
\end{equation}
provides a viable model for the four-parton correlation term in (\ref{eq8a}). If desired, taking more complicated linear combinations of the functions (\ref{ansatz}) one could construct functions that are free of any large logarithms in the high-energy limit. For instance, the combination $F_{11}-2F_{12}+F_{22}\to-i\pi^5/8$ tends to a constant for $L\to\infty$.

The result (\ref{testfun}) proves that the high-energy Regge limit cannot be used to exclude the presence of the four-parton term proportional to $F$ in (\ref{eq8a}). Only an actual three-loop calculation of a four-parton scattering amplitude beyond the planar limit can show whether or not this term is present in the anomalous dimension. The fate of the dipole conjecture depends on the result of such a calculation. First steps toward the computation of the three-loop four-gluon amplitude in ${\cal N}=4$ super Yang-Mills theory were made in \cite{Bern:2008pv}.

\section{Diagrammatic analysis at four loops}
\label{a.3}

\begin{figure}[t]
\begin{center}
\includegraphics[width=0.55\textwidth]{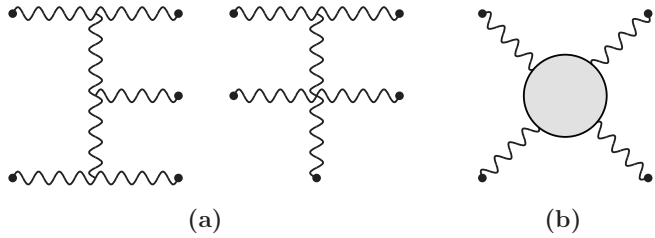}
\parbox{14.5cm}{\caption{\label{4loop} 
Four-loop connected webs contributing to the soft anomalous dimension. The dots represent color generators, which appear when the gluons are attached to the Wilson lines representing the external partons. The webs in (b) consist of the color-symmetrized parts of one-loop quark, gluon, and ghost diagrams with four external gluons.}}
\end{center}
\end{figure}

Contributions of $\ord(\as^4)$ to the soft anomalous dimension (\ref{eq8}) can arise from the gluon webs shown in Figure~\ref{4loop}. At this order there exist additional two- and three-gluon webs containing insertions of self-energies or vertex functions, which however give rise to the 
same color structures appearing already at two- and three-loop order. They do not give rise to new structures in the anomalous dimension. The two five-gluon webs in Figure~\ref{4loop} (a) are described by the same color structure $f^{adx}f^{bcy}f^{exy}\,\T^a_i\T^b_j\T^c_k\T^d_l\T^e_m$. If two or more color generators act on the same parton, the product of generators with the same parton index can be decomposed into symmetric 
and antisymmetric parts. The Lie-algebra relations
\begin{equation}
\begin{aligned}
   \big[ \T^a , \T^b \big] &= i f^{abc}\T^c \,, \qquad
    f^{abc} f^{abd} = C_A\,\delta^{cd} \,, \\
   \mbox{tr}_{\rm adj}(\T^a\T^b\T^c) 
   &= i f^{ade} f^{beg} f^{cgd} = \frac{i C_A}{2}\,f^{abc}
\end{aligned}   
\end{equation}
can be used to reduce the antisymmetric products to structures containing fewer generators. It is therefore sufficient to consider symmetrized products of color generators only. The four-gluon
webs in Figure~\ref{4loop} (b) instead are proportional to the traces 
\begin{equation}
   d_R^{a_1 a_2\dots a_n} 
   = \mbox{tr}\big[ \left( \T_R^{a_1}\T_R^{a_2}\dots\T_R^{a_n} \right)_+ \big]
\end{equation}
of symmetrized products of generators in the representation $R$. Contracting them with the color generators of the external partons gives rise to higher Casimir invariants. As discussed in \cite{Becher:2009qa}, it is sufficient to consider traces of generators in the fundamental representation, since $d_R^{a_1 a_2\dots a_n}=I_n(R)\,d_F^{a_1 a_2\dots a_n}$ with a representation-dependent index $I_n(R)$. The coefficient functions $g_i$ and $G_1$ in relation (\ref{54}) below contain two terms of the form $g_i(\alpha_s)=n_f\,g_i^F(\alpha_s)+I_4(A)\,g_i^A(\alpha_s)$, and similarly for $G_1$. It follows that we need to consider the two color structures
\begin{equation}\label{50a}
\begin{aligned}
   &\mbox{(a)} \quad &
   {\cal T}_{ijklm} &= f^{adx} f^{bcy} f^{exy}
    \left( \T^a_i\T^b_j\T^c_k\T^d_l\T^e_m \right)_+ , \\
   &\mbox{(b)} \quad &
   {\cal D}_{ijkl} &= d_F^{abcd} \left( \T^a_i \T^b_j \T^c_k \T^d_l \right)_+ ,
\end{aligned}
\end{equation}
which we will analyze separately in the next two sections. 

In our discussion in this section we will {\em not\/} make use of the constraint derived by considering two-particle collinear limits, since the validity of collinear factorization has not yet been proved to all orders in perturbation theory. We first only use the constraints implied by non-abelian exponentiation and soft-collinear factorization. The additional simplifications implied when one considers two-particle collinear limits will be discussed in Section~\ref{sec:2p_collinear}. Note that without the collinear constraint there can also be an additional contribution to the soft anomalous dimension (\ref{eq8a}) starting at three-loop order, which reads \cite{Becher:2009qa}
\begin{equation}\label{dG3}
   {\bf\Delta\Gamma}_3 = \bar f_2(\as) \sum_{(i,j)}\,{\cal T}_{iijj}
   = - \bar f_2(\as) \sum_{(i,j,k)}\,{\cal T}_{iijk} \,.
\end{equation}
The subscript 3 indicates that this terms is at least of $\ord{(\as^3)}$.

\subsection{Four-gluon webs}

We begin by considering terms involving the color structure (b) in (\ref{50a}), which are associated with the four-gluon web shown in Figure~\ref{4loop} (b). The most general contribution to ${{\bf\Gamma}_s}$ compatible with the symmetry properties of this structure reads
\begin{equation}\label{54}
\begin{aligned}
   {\bf\Delta\Gamma}_{s4}^{(b)} 
   &= \sum_{(i,j)}\,\big[ {\cal D}_{iijj}\,g_1(\as) + {\cal D}_{iiij}\,g_2(\as) \big]\,\beta_{ij}
    + \sum_{(i,j,k)} {\cal D}_{ijkk}\,g_3(\as)\,\beta_{ij} \\ 
   &\quad\mbox{}+ \sum_i\,{\cal D}_{iiii}\,g_4(\as)
    + \sum_{(i,j)}\,{\cal D}_{iijj}\,g_5(\as) 
    + \sum_{(i,j,k,l)} {\cal D}_{ijkl}\,G_1(\beta_{ijkl},\beta_{iklj},\beta_{iljk}) \,,
\end{aligned}
\end{equation}
where we have used the relation
\begin{equation}\label{colorcons}
   \sum_i \T_i^a = 0 \quad \mbox{when acting on color-singlet states,}
\end{equation} 
which is implied by color conservation, to simplify the result. The terms in the first line, which are linear in the cusp angles $\beta_{ij}$, have already been analyzed in \cite{Becher:2009qa}, where it was shown that the constraint from soft-collinear factorization implies the relations
\begin{equation}
   g_2(\as) = 2 g_1(\as) \,, \qquad g_3(\as) = g_1(\as) \,.
\end{equation}
The remaining structures shown in the second line were not considered before. Starting at the four-parton level a non-trivial function of conformal cross ratios can appear. The term proportional to $g_4(\as)$ contributes to the single-particle anomalous dimension $\gamma_s^i$ in (\ref{eq8}). Its coefficient ${\cal D}_{iiii}=C_4(F,R_i)$ is given in terms of a higher Casimir invariant \cite{Becher:2009qa}, depending on the representation of parton $i$. 

The symmetries of the color structure ${\cal D}_{ijkl}$ imply that the function $G_1(\beta_{ijkl},\beta_{iklj},\beta_{iljk})$ must be invariant under any cyclic or anti-cyclic permutation of its arguments. This is the reason why we have written it as a function of the three conformal cross ratios $\beta_{ijkl}$, $\beta_{iklj}$, $\beta_{iljk}$, even though according to (\ref{lindep}) only two of them are linearly independent. The simplest example of such a function, which has maximal transcendentality at four loops ($\tau=6$) and in addition vanishes in all relevant two-particle collinear limits, is $G_1(x,y,z)=x^2 y^2 z^2$ \cite{Dixon:2009ur}, but this simple form is not compatible with the constraints implied by the Regge limit. In analogy with our discussion in Section~\ref{a.2}, it would however not be difficult to construct more complicated functions involving products of logarithms and dilogarithms, which satisfy both the collinear and Regge constraints.

\subsection{Five-gluon webs}

Next we analyze possible contributions to the soft anomalous dimension arising from the color structure (a) in (\ref{50a}). It has the symmetry properties
\begin{equation}\label{101}
   {\cal T}_{ijklm} = - {\cal T}_{ikjlm} = - {\cal T}_{ljkim} 
   = - {\cal T}_{jilkm} \,,
\end{equation}
which allow us to move any one of the first four indices to first place. Note that the fifth index is special and cannot be moved. In the following we will consider attachments of this structure to different numbers of Wilson lines representing the external particles.

The symmetry properties (\ref{101}) imply that the color structure ${\cal T}_{ijklm}$ vanishes whenever there are less than three different parton indices involved. For the case of three different indices $i,j,k$, the symmetry properties allow us to reduce all possible structures to ${\cal T}_{iijki}$ and ${\cal T}_{iikjj}$, where the first one is antisymmetric in $j,k$. The Jacobi identity implies that the second structure is asymmetric in $i,j$. To see this, note that 
\begin{equation}\label{tauasym}
\begin{aligned}
   {\cal T}_{iikjj} + {\cal T}_{jjkii}
   &= f^{adx} \left( f^{bcy} f^{exy} + f^{cey} f^{bxy} \right)
    \left( \T_i^a \T_i^b\right)_+ \T_k^c \left( \T_j^d \T_j^e\right)_+  \\    
   &= - f^{adx} f^{bey} f^{xcy} 
    \left( \T_i^a \T_i^b\right)_+ \T_k^c \left( \T_j^d \T_j^e\right)_+ 
    = 0 \,.
\end{aligned}
\end{equation} 
The last equation follows because the product of color generators is invariant under the combined exchange $a\leftrightarrow b$ and $d\leftrightarrow e$, whereas the product of structure constants is odd under this exchange. For the case of four different indices $i,j,k,l$, the symmetry properties imply that all structures can be reduced to ${\cal T}_{iijkl}$ and ${\cal T}_{ijkli}$, both of which are antisymmetric in $j,k$. Up to this point, the most general contribution to the soft anomalous dimension has the form
\begin{equation}\label{103}
\begin{aligned}
   {\bf\Delta\Gamma}_{s4}^{(a)\,{\rm part 1}} 
   &= \sum_{(i,j,k)} \Big[ {\cal T}_{iijki}\,g_6(\as)\,\beta_{ij} 
    + {\cal T}_{iikjj}\,g_7(\as)\,\beta_{ik} \Big] \\
   &\quad\mbox{}+ \sum_{(i,j,k,l)}\!\bigg\{ 
    {\cal T}_{iijkl} \Big[ g_8(\as)\,\beta_{ij} + g_9(\as)\,\beta_{il} 
    + g_{10}(\as)\,\beta_{jl} + G_2(\beta_{ijkl},\beta_{iklj}) \Big] \\
   &\hspace{16.8mm}\mbox{}+ {\cal T}_{ijkli} \Big[ g_{11}(\as)\,\beta_{ij} 
    + g_{12}(\as)\,\beta_{il} + g_{13}(\as)\,\beta_{jl} 
    + G_3(\beta_{ijkl},\beta_{iklj}) \Big] \bigg\} \,.
\end{aligned}
\end{equation}
The $G_n$ functions depend on two linearly independent cross ratios, for which we have chosen $\beta_{ijkl}$ and $\beta_{iklj}$. Note that the symmetry properties of the color structure ${\cal T}_{ijklm}$ imply that no constant terms, independent of cusp angles and conformal cross ratios, can appear in this result.  

Color conservation can be used to simplify the above expression by performing the sum over those parton indices not involved in the definition of the various cusp angles. To this end, we move the corresponding color generator to the right in the symmetrized product ${\cal T}_{ijklm}$ in (\ref{50a}) and then apply the property (\ref{colorcons}). Renaming some of the summation indices, we then obtain
\begin{equation}\label{G5a}
   {\bf\Delta\Gamma}_{s4}^{(a)\,{\rm part 1}} 
   = \sum_{(i,j,k)} {\cal T}_{iikjj}\,\bar g_7(\as)\,\beta_{ik}
    + \sum_{(i,j,k,l)}\!\Big[ {\cal T}_{iijkl}\,G_2(\beta_{ijkl},\beta_{iklj})  
    + {\cal T}_{ijkli}\,G_3(\beta_{ijkl},\beta_{iklj}) \Big] \,,
\end{equation}
where $\bar g_7\equiv g_7-g_8+g_{10}+g_{11}-g_{13}$. The soft-collinear factorization constraint (\ref{eq9}) does not imply a non-trivial condition on the coefficient function $\bar g_7$. After summing over free parton indices and using the symmetry property (\ref{tauasym}), we obtain
\begin{equation}
   \sum_{j\ne i}\,\frac{\partial{\bf\Delta\Gamma}_{s4}^{(a)\,{\rm part 1}}}{\partial\beta_{ij}} 
   = \sum_{(j\ne i,k\ne i)}\!\big( {\cal T}_{iijkk} + {\cal T}_{jjikk} \big)\,\bar g_7(\as) 
   = 0 \,.
\end{equation}
It follows that ${\bf\Delta\Gamma}_{s4}^{(a)\,{\rm part 1}}$ is a solution to the homogeneous equation, and it should therefore be possible to express it entirely in terms of conformal cross ratios. This is indeed possible when one exploits that the three-parton and four-parton terms in (\ref{103}) can be related to each other using color conservation, as is evident from the structure of (\ref{G5a}). There are several equivalent ways to write the final answer. For example, we find \begin{equation}
   {\bf\Delta\Gamma}_{s4}^{(a)\,{\rm part 1}} 
   = \sum_{(i,j,k,l)}\!\Big\{ 
    {\cal T}_{iijkl}\,\Big[ G_2(\beta_{ijkl},\beta_{iklj}) 
    + \frac12\,\bar g_7(\as)\,\beta_{iklj} \Big] 
    + {\cal T}_{ijkli}\,G_3(\beta_{ijkl},\beta_{iklj}) \Big\} \,,
\end{equation}
which implies that the effect of $\bar g_7(\as)$ can be absorbed into a redefinition of the function $G_2$. We will simply drop this contribution from now on, but for simplicity we will refrain from renaming the function $G_2$. 

It remains to consider the case of five different parton indices $i,j,k,l,m$, for which it is straightforward to show that all terms linear in cusp angles vanish by color conservation. This only leaves the possibility of non-linear functions of conformal cross ratios. Given five different parton indices, there are five subgroups of four parton indices, such as $i,j,k,l$. For each subgroup there exist only two linearly independent conformal cross ratios, in analogy with our discussion around (\ref{lindep}). However, among the remaining ten cross ratios there exist five additional linear relations, so that only five linearly independent cross ratios remain. They can be chosen as $\beta_{ijkl}$, $\beta_{iklj}$, $\beta_{ijkm}$, $\beta_{ikmj}$, and $\beta_{ijml}$. Thus
\begin{equation}\label{G5b}
   {\bf\Delta\Gamma}_{s4}^{(a)\,{\rm part 2}} 
   = \sum_{(i,j,k,l,m)}\!{\cal T}_{ijklm}\,
   G_4(\beta_{ijkl},\beta_{iklj},\beta_{ijkm},\beta_{ikmj},\beta_{ijml}) \,.
\end{equation}
The complete result is given by the sum of the two contributions in (\ref{G5a}) and (\ref{G5b}).

\subsection{\boldmath Combined four-loop result and implications for $\Gamma_{\rm cusp}^i$}

We can now combine the above results and work out the form of the new contributions to the anomalous dimension (\ref{eq8a}) arising at four-loop order. To this end, we have to convert the cusp angles $\beta_{ij}$ into logarithms of kinematic invariants using
\begin{equation}
   \beta_{ij} = - \ln\frac{\mu^2}{-s_{ij}} + \ln\frac{\mu^2}{-p_i^2} + \ln\frac{\mu^2}{-p_j^2} \,.
\end{equation}
The conformal cross ratios $\beta_{ijkl}$ have already been expressed in terms of $s_{ij}$ variables in the second relation of (\ref{betadef}). When the above expression for $\beta_{ij}$ is inserted in (\ref{54}), the terms proportional to the collinear logarithms $\ln\mu^2/(-p_i^2)$ multiply color-singlet structures, which can be cancelled against corresponding terms in the collinear anomalous dimension ${\bf\Gamma}_c$ in (\ref{Gammac}). We also include the single-particle contribution proportional to $g_4(\as)$ in (\ref{54}) into the anomalous dimension $\gamma_s^i$ in (\ref{eq8}), which becomes part of $\gamma^i$ in (\ref{eq8a}). For the remaining contributions, we obtain
\begin{eqnarray}\label{dG4}
   {\bf\Delta\Gamma}_4 
   &=& - g_1(\as) \bigg[ \sum_{(i,j)}\,\big( {\cal D}_{iijj} + 2 {\cal D}_{iiij} \big)
    \ln\frac{\mu^2}{-s_{ij}}
    + \sum_{(i,j,k)} {\cal D}_{ijkk}\,\ln\frac{\mu^2}{-s_{ij}} \bigg] 
    + \sum_{(i,j)}\,{\cal D}_{iijj}\,g_5(\as) \nonumber\\ 
   &&\mbox{}+ \sum_{(i,j,k,l)}\!\Big[ 
    {\cal D}_{ijkl}\,G_1(\beta_{ijkl},\beta_{iklj},\beta_{iljk}) 
    + {\cal T}_{iijkl}\,G_2(\beta_{ijkl},\beta_{iklj}) 
    + {\cal T}_{ijkli}\,G_3(\beta_{ijkl},\beta_{iklj}) \Big] \nonumber\\
   &&\mbox{}+ \sum_{(i,j,k,l,m)}\!{\cal T}_{ijklm}\,
    G_4(\beta_{ijkl},\beta_{iklj},\beta_{ijkm},\beta_{ikmj},\beta_{ijml}) \,.
\end{eqnarray}

In the special case of two external particles in the color representation $R_i$, the result for the anomalous dimension simplifies considerably. From (\ref{eq8a}), (\ref{dG3}), and (\ref{dG4}), we then obtain
\begin{equation}
   {\bf\Gamma} 
   = - \Big[ C_i\,\gamma_{\rm cusp}(\as) - 2 g_1(\as)\,C_4(F,R_i) \Big]\,\ln\frac{\mu^2}{-s_{12}}
    + 2\gamma^i(\as) + 2 C_4(F,R_i) \,g_5(\as) \,.
\end{equation}
The coefficient of the logarithm is identified with the cusp anomalous dimension of the particle. In generalization with (\ref{casimir}), we find
\begin{equation}\label{casimir4}
   \Gamma_{\rm cusp}^i(\as) 
   = C_i\,\gamma_{\rm cusp}(\as) - 2 g_1(\as)\,C_4(F,R_i) + \ord(\as^5) \,.
\end{equation}
The four-loop term proportional to the higher Casimir invariant $C_4(F,R_i)$ would violate the Casimir scaling relation $\Gamma_{\rm cusp}^q(\as)/C_F=\Gamma_{\rm cusp}^g(\as)/C_A$, provided that $g_1(\as)\ne 0$. The four-loop cusp anomalous dimension is known for ${\cal N}=4$ super Yang-Mills theory in the planar limit \cite{Bern:2006ew}. However, as shown in \cite{Becher:2009qa} the higher Casimir contributions are subleading for large $N_c$. These structures are therefore not visible in the planar limit of $SU(N_c)$ gauge theory.

\subsection{Two-particle collinear limits}
\label{sec:2p_collinear}

Additional constraints on the anomalous dimension can be derived by considering the limit where the momenta of two or more external partons become collinear. Assuming that $n$-parton scattering amplitudes in the collinear limit factorize into $(n-1)$-parton scattering amplitudes times universal splitting amplitudes \cite{Berends:1988zn,Mangano:1990by,Bern:1995ix}, the anomalous dimension of the splitting amplitudes must satisfy the relation \cite{Becher:2009qa}
\begin{equation}\label{GamSP}
   {\bf\Gamma}_{\rm Sp}(\{p_1,p_2\},\mu)
   = {\bf\Gamma}(\{p_1,\dots,p_n\},\mu) - {\bf\Gamma}(\{P,p_3,\dots,p_n\},\mu) 
    \big|_{\T_P\to\T_1+\T_2} \,,
\end{equation}
where $P=p_1+p_2$ denotes the sum of the momenta of the two collinear partons. A non-trivial constraint arises from the fact that the anomalous dimension of the $P\to 1+2$ splitting amplitude can only depend on the momenta and color matrices of the two collinear partons.

While this constraint is satisfied for the anomalous dimension in (\ref{eq8a}) as long as the function $F$ vanishes in all two-particle collinear limits, additional constraints arise for the correction terms in (\ref{dG3}) and (\ref{dG4}). The corresponding contributions to the anomalous dimension of the splitting amplitudes read
\begin{equation}\label{eq37}
\begin{aligned}
   {\bf\Delta\Gamma}_{\rm Sp}(\{p_1,p_2\},\mu)
   &= \bar f_2(\as)\,\bigg( 2{\cal T}_{1122} - 4\sum_{i\ne 1,2}\,{\cal T}_{12ii} \bigg)
    + g_5(\as)\,\bigg( 2{\cal D}_{1122} - 4\sum_{i\ne 1,2}\,{\cal D}_{12ii} \bigg) \\
   &\quad\mbox{}- 2g_1(\as)\,\bigg[ \bigg( {\cal D}_{1122} + {\cal D}_{1112} + {\cal D}_{1222} 
    + \sum_{i\ne 1,2}\,{\cal D}_{12ii} \bigg)\,\ln\frac{\mu^2}{-s_{12}} 
    + \dots \bigg] \,,
\end{aligned}
\end{equation}
where for the contribution proportional to $g_1$ we only show the terms proportional to $\ln[\mu^2/(-s_{12})]$ for simplicity \cite{Becher:2009qa}. In addition, we must require that all functions of conformal cross ratios vanish in the relevant two-particle collinear limits. All three terms shown above contain structures that are quadratic in the color generators of partons not involved in the splitting process, which is in conflict with the assumption of collinear factorization. We thus conclude that 
\begin{equation}\label{cols}
   \bar f_2(\as) = 0 \,, \qquad
   g_1(\as) = 0 \,, \qquad
   g_5(\as) = 0 \,,
\end{equation}
where the first two relations were already derived in \cite{Becher:2009qa}. It follows that ${\bf\Delta\Gamma}_3=0$ in (\ref{dG3}), and in ${\bf\Delta\Gamma}_4$ in (\ref{dG4}) the terms shown in the first line vanish, such that only the functions of conformal cross ratios remain:
\begin{eqnarray}\label{dG4fin}
   {\bf\Delta\Gamma}_4 
   &=& \sum_{(i,j,k,l)}\!\Big[ 
    {\cal D}_{ijkl}\,G_1(\beta_{ijkl},\beta_{iklj},\beta_{iljk}) 
    + {\cal T}_{iijkl}\,G_2(\beta_{ijkl},\beta_{iklj}) 
    + {\cal T}_{ijkli}\,G_3(\beta_{ijkl},\beta_{iklj}) \Big] \nonumber\\
   &&\mbox{}+ \sum_{(i,j,k,l,m)}\!{\cal T}_{ijklm}\,
    G_4(\beta_{ijkl},\beta_{iklj},\beta_{ijkm},\beta_{ikmj},\beta_{ijml}) \,.
\end{eqnarray}
The functions $G_i$ must vanish in all relevant two-particle collinear limits. The functional dependence of the five-parton structure $G_4$ can in addition be constrained by considering multi-particle collinear limits, in which more than two parton momenta become collinear. This is briefly discussed in Appendix~\ref{ap.2}. In analogy with our treatment in Section~\ref{a.2}, it would most likely not be difficult to construct explicit examples of $G_n$ functions that are consistent with all constraints from collinear limits and in addition are free of leading logarithms in the Regge limit. 

The most interesting consequence of the relations (\ref{cols}) is that the extra term proportional to $C_4(F,R_i)$ to the cusp anomalous dimension in (\ref{casimir4}) vanishes, so that Casimir scaling still holds at $\ord{(\as^4)}$ \cite{Becher:2009qa}. This prediction is highly non-trivial in view of the expectation that Casimir scaling should not hold non-perturbatively, at least not for the finite parts of Wilson-loop expectation values. A long time ago, Frenkel and Taylor argued that Casimir scaling would be inconsistent with expectations about the area law for matrix elements of Wilson loops giving rise to confinement \cite{Frenkel:1984pz}. More recently, investigations of high-spin operators in string theory using the AdS/CFT correspondence have found a strong-coupling behavior that is in conflict with Casimir scaling \cite{Armoni:2006ux,Alday:2007hr,Alday:2007mf}.

When deriving the constraints from two-particle collinear limits in the previous section, we have implicitly assumed that the number of external partons is $n\ge 4$. The cases $n=2$ and $n=3$ are somewhat special, in the sense that for $n=2$ the two-particle collinear limit does not apply, while for $n=3$ the color-conservation relation (\ref{colorcons}) can be used to rewrite the third color generator in terms of the color generators of the two particles whose momenta become collinear. As a result, the conclusions (\ref{cols}) can no longer be drawn. We emphasize, however, the important point that the collinear anomalous dimension in (\ref{Gammac}) has the same form irrespective of the number of partons involved in the scattering process. For $n\ge 4$, our arguments imply that the coefficient $\Gamma_{\rm cusp}^i(\as)$ entering this expression obeys Casimir scaling, and hence it follows that $g_1(\as)=0$ in (\ref{casimir4}) also for $n=2,3$. In addition, the process independence of the splitting amplitudes excludes the possibility of having extra contributions for the special case where $n=3$, and hence we can conclude from (\ref{eq37}) that $\bar f_2(\as)=0$ and $g_5(\as)=0$ also in this case. Finally, the last relation in (\ref{dG3}) implies that the contribution proportional to $\bar f_2$ is simply absent for $n=2$. On the other hand, we cannot exclude the possibility that the term proportional to $g_5$ in (\ref{dG4}) is present in this particular case. Strictly speaking, one should therefore add the term $2C_4(F,R_1)\,g_5(\as)\,\delta_{n2}$ to the right-hand side of (\ref{dG4fin}), even though the presence of such an exceptional contribution would seem strange.

The validity of collinear factorization has recently been scrutinized \cite{Catani:2011st,Forshaw:2012bi}. Our analysis in this section relies on the assumption that this property is true to sufficiently high orders in perturbation theory, and this then implies relation (\ref{GamSP}), which imposes an important constraint on the structure of the anomalous dimension in higher orders. An all-order proof of collinear factorization for leading-color amplitudes has been given in \cite{Kosower:1999xi}. The authors of \cite{Catani:2011st} have argued that one should distinguish between a time-like collinear limit, in which the two collinear partons are both either in the initial or the final state, and a space-like collinear limit, in which one particle is in the initial state and the other one in the final state. They have considered the known dipole structure of IR divergences and derived the corresponding structure of the anomalous dimensions of the splitting amplitudes. They have found that collinear factorization is satisfied for a time-like, but not for a space-like collinear limit. The explanation for this fact has been investigated further in \cite{Forshaw:2012bi}. Briefly, factorization in the time-like collinear limit is guaranteed by color coherence. The latter is at work in the space-like collinear limit as well, but there factorization is broken by the non-commutativity of Coulomb/Glauber gluon exchanges with other soft exchanges.

For our purposes, however, it is sufficient to only consider time-like collinear limits. Then there is no problem with the constraint (\ref{GamSP}), and it can be used to derive the desired conditions on the anomalous dimension. Note that even though $\bf\Gamma$ controls the IR divergences of scattering amplitudes, at the same time it controls the UV divergences of $n$-jet operators in SCET. UV singularities are independent of external states and do not care whether we consider matrix elements with particles in the initial or final state.

\subsection{\boldmath Large-$N_c$ limit}

It is interesting to explore how the new structures arising at four-loop order and compiled in (\ref{dG4fin}) behave in the large-$N_c$ limit of $SU(N_c)$ gauge theory. Since several results at three- and four-loop order are known in the $N_c\to\infty$ (or ``planar'') limit for ${\cal N}=4$ super Yang-Mills theory, one might hope that they can be used to constrain or even determine some of the functions of conformal cross ratios. Unfortunately, the three-loop four-particle term proportional to $F$ in (\ref{eq8a}) is suppressed relative to the leading color-dipole term and hence does not give a contribution in the planar limit \cite{Becher:2009qa}. We will now show that the same is true for the new structures arising at four-loop order.

In order to illustrate this fact, we follow \cite{Becher:2009qa} and consider the effect of the color structures in (\ref{dG4fin}) acting on color traces of the type ${\rm tr}(t^{a_1}\dots t^{a_n})$, which represent a basis of leading color structures of $n$-particle gluonic amplitudes. In this way, we find that
\begin{equation}\label{leadcol}
\begin{aligned}
   {\cal D}_{ijkl} &= {\cal O}(N_c) \,, &\qquad
    {\cal T}_{iijkl} &= {\cal O}(N_c^2) \,, \\
   {\cal T}_{ijkli} &= {\cal O}(N_c^2), &\qquad
    {\cal T}_{ijklm} &= {\cal O}(N_c^2) \,,
\end{aligned}
\end{equation}
where all indices are assumed to be different. The leading color structures at $n$-loop order are of the form $N_c^n\,{\rm tr}(t^{a_1}\dots t^{a_n})$. The above structures appear first at four-loop order in (\ref{dG4fin}) and therefore are suppressed compared to the leading term by at least a factor $1/N_c^2$. Similarly, the three-loop structure ${\cal T}_{ijkl}$ in (\ref{eq8a}) is of ${\cal O}(N_c)$ \cite{Becher:2009qa}, which is $1/N_c^2$-suppressed relative to the leading structure at three-loop order.

\section{Conclusions}
\label{a.4}

IR divergences of massless gauge-theory scattering amplitudes can be removed by means of a multiplicative renormalization factor $\Z$, which in turn is determined by the anomalous dimension $\bf\Gamma$ of $n$-jet operators in SCET. The structure of this anomalous dimension is strongly constrained by color conservation, soft-collinear factorization, non-abelian exponentiation, two-particle (and multi-particle) collinear limits, and the high-energy Regge limit. When these constraints are taken into account, it is found that up to three-loop order the anomalous dimension takes the simple form (\ref{eq8a}). The terms in the first line of this equation, which start at one-loop order, describe pairwise color-dipole correlations among the different particles. The dipole conjecture states that these terms are all there is to all orders in perturbation theory \cite{Becher:2009cu,Gardi:2009qi,Becher:2009qa}. At ${\cal O}(\as^3)$, the only possible exception is an additional structure inducing correlations among four external particles, shown in the second line of (\ref{eq8a}). So far, no example of a function $F(x,y)$ of conformal cross ratios consistent with the constraints from collinear limits and the high-energy Regge limit had been constructed in the literature, and it had been speculated whether these constraints combined might even exclude the existence of such a function~\cite{DelDuca:2011ae}.

We have demonstrated in Section~\ref{a.2} that this hope was too optimistic. We have constructed a class of rather simple functions with maximum transcendentality, which satisfy all constraints. This is an existence proof that non-trivial functions of conformal cross ratio, arising at the level of three loops and higher, are possible and not excluded by any known constraints. Indeed, our analysis shows that while the Regge limit constrains the functional form of these functions, it does not imply any new structural constraints on the anomalous dimension beyond those derived from soft-collinear factorization, non-abelian exponentiation, and collinear limits. In Section~\ref{a.3} we have extended the analysis of the anomalous dimension to four-loop order, finding further modifications of the dipole formula that once again are governed by functions of conformal cross ratios. The most general structure is shown in (\ref{dG4fin}). It involves three new types of four-particle correlations and a single type of five-particle correlation. While the functional dependence of these structures can be constrained by considering two-particle (and multi-particle) collinear limits and the high-energy behavior, their presence cannot be excluded, and following the lines of Section~\ref{a.2} it would not be difficult to construct examples of functions consistent with all constraints.

In light of these findings, it is not unlikely that the dipole conjecture must be revised and that all or some of the additional structures will indeed turn out to be non-zero. Based on the observations made in the previous sections, we are led to propose a weaker form of the conjecture, which states that violations of the dipole formula, if they exist, share three properties: 
\begin{itemize}
\item
they involve correlations of at least four partons;
\item
they are given in terms of functions of conformal cross ratios, which vanish in all relevant collinear limits and are free of leading logarithms in the Regge limit;
\item
at each order in $\as$, they are suppressed relative to the contributions contained in the dipole formula by at least a factor of $1/N_c^2$.
\end{itemize}
The last statement implies that the dipole formula is exact (in perturbation theory) in the planar limit for $SU(N_c)$ gauge theory. The first statement implies that the dipole formula holds to all orders in perturbation theory for the case of scattering amplitudes involving two or three partons. It then follows that the cusp anomalous dimension obeys Casimir scaling: 
\begin{equation}
   \Gamma_{\rm cusp}^i(\as) = C_i\,\gamma_{\rm cusp}(\as) 
   \quad \mbox{(perturbatively).}
\end{equation}
While this relation is known to be true to ${\cal O}(\as^3)$ by direct calculation \cite{Moch:2004pa}, it is not expected to hold non-perturbatively \cite{Frenkel:1984pz,Armoni:2006ux,Alday:2007hr,Alday:2007mf}. The weak dipole conjecture could be violated if some of our basic assumptions, such as the validity of collinear factorization, were to fail starting at some order in perturbation theory. In (\ref{dG4}) and (\ref{casimir4}) we have presented expressions for the anomalous dimension in general, and the cusp anomalous dimension in particular, which can be derived without making use of the collinear constraint. It would be very interesting to check the prediction of Casimir scaling of the cusp anomalous dimension at four-loop order by an explicit calculation.

\subsubsection*{Acknowledgements}

We are grateful to Thomas Becher and Lorenzo Magnea for useful comments and suggestions. The research of M.N.\ is supported by the Advanced Grant EFT4LHC of the European Research Council (ERC), grant NE 398/3-1 of the German Research Foundation (DFG), grants 05H09UME and 05H12UME of the German Federal Ministry for Education and Research (BMBF), and the Rhineland-Palatinate Research Center {\em Elementary Forces and Mathematical Foundations}. The work of V.A.\ was supported in part by the DFG Graduate Training Center GRK 1581. L.V.\ acknowledges the Alexander von Humboldt Foundation for support.

\appendix
\section{Four-loop expression for $\Z$}
\label{ap.1}
\renewcommand{\theequation}{A\arabic{equation}}
\setcounter{equation}{0}

Given a UV renormalized, on-shell $n$-parton scattering amplitude $|{\cal M}_n(\ep,\{\underline p\})\rangle$ with IR divergences regularized in $d=4-2\ep$ dimensions, one obtains the finite amplitude $|{\cal M}_n(\{\underline p\},\mu)\rangle$, in which all IR are subtracted in a minimal way, from the relation \cite{Becher:2009cu}
\begin{equation}
   |{\cal M}_n(\{\underline p\},\mu)\rangle 
   = \lim_{\ep\to 0} \Z^{-1}(\ep,\{\underline p\},\mu)\,|{\cal M}_n(\ep,\{\underline p\})\rangle \,.
\end{equation}
The $\Z$ factor is related to the anomalous dimension $\bf\Gamma$ studied in the present paper by 
\begin{equation}
   {\bf\Gamma}(\{\underline p\},\mu) 
   = - \Z^{-1}(\ep,\{\underline p\},\mu)\,\frac{d}{d\ln\mu}\,\Z(\ep,\{\underline p\},\mu) \,.
\end{equation}
A formal solution to this equation was derived in \cite{Becher:2009qa}, where the perturbative expansion of $\ln\Z$ was constructed to ${\cal O}(\as^3)$. Including also the next term in the series, we obtain
\begin{eqnarray}
   \ln{\bf Z} &=& \frac{\as}{4\pi}
    \left( \frac{\Gamma_0'}{4\epsilon^2} + \frac{{\bf\Gamma}_0}{2\epsilon} \right)
    + \left( \frac{\as}{4\pi} \right)^2 \left( - \frac{3\beta_0 \Gamma_0'}{16\epsilon^3}
    + \frac{\Gamma_1'-4\beta_0 {\bf\Gamma}_0}{16\epsilon^2}
    + \frac{{\bf\Gamma}_1}{4\epsilon} \right) \nonumber\\
   &&\mbox{}+ \left( \frac{\as}{4\pi} \right)^3
    \bigg( \frac{11\beta_0^2\Gamma_0'}{72\epsilon^4}
    - \frac{5\beta_0\Gamma_1'+8\beta_1\Gamma_0'-12\beta_0^2{\bf\Gamma}_0}{72\epsilon^3} 
    + \frac{\Gamma_2'-6\beta_0{\bf\Gamma}_1-6\beta_1{\bf\Gamma}_0}{36\epsilon^2}
    + \frac{{\bf\Gamma}_2}{6\epsilon} \bigg) \nonumber\\
   &&\mbox{}+ \left( \frac{\as}{4\pi} \right)^4
    \bigg( - \frac{25\beta_0^3\Gamma_0'}{192\epsilon^5}
    + \frac{13\beta_0^2\Gamma_1'+40\beta_0\beta_1\Gamma_0'-24\beta_0^3{\bf\Gamma}_0}{192\epsilon^4} \\ 
   &&\hspace{20.5mm}\mbox{}- \frac{7\beta_0\Gamma_2'+9\beta_1\Gamma_1'+15\beta_2\Gamma_0'
    -24\beta_0^2{\bf\Gamma}_1-48\beta_0\beta_1{\bf\Gamma}_0}{192\epsilon^3} \nonumber\\
   &&\hspace{20.5mm}\mbox{}+ 
    \frac{\Gamma_3'-8\beta_0{\bf\Gamma}_2-8\beta_1{\bf\Gamma}_1-8\beta_2{\bf\Gamma}_0}{64\epsilon^2}
    + \frac{{\bf\Gamma}_3}{8\epsilon} \bigg) + {\cal O}(\as^5) \,, \nonumber
\end{eqnarray}
where we have expanded the anomalous dimension and $\beta$-function as 
\begin{equation}
   {\bf\Gamma}(\as) = \sum_{n=0}^{\infty} {\bf\Gamma}_n \left( \frac{\as}{4\pi} \right)^{n+1} ,
    \qquad
   \beta(\as) = -2\as \sum_{n=0}^{\infty} \beta_n \left( \frac{\as}{4\pi} \right)^{n+1} ,
\end{equation}
and similarly for the function $\Gamma'(\as)=-\gamma_{\rm cusp}(\as)\,\sum_i C_i$.

\section{Multi-particle collinear limits}
\label{ap.2}
\renewcommand{\theequation}{B\arabic{equation}}
\setcounter{equation}{0}

Relation (\ref{GamSP}), which is a consequence of collinear factorization, can be extended to the case where more than two particle momenta become collinear. Given an $n$-parton scattering process in the limit in which $m$ partons become collinear, the amplitude factorizes into an $(n-m+1)$-parton amplitude times a process-independent splitting function, which involves only the $m$ collinear partons (see e.g.~\cite{Catani:2003vu} for the case $m=3$). More explicitly, assigning momenta $p_1=z_1 P$, $p_2=z_2 P$, \dots, $p_m=z_m P$ with $z_1+z_2+\dots+z_m=1$, 
the amplitude factorizes as
\begin{equation}\label{30}
   |{\cal M}_n(\{p_1,\dots,p_m,p_{m+1},\dots, p_n\})\rangle 
   = {\bf Sp}(\{p_1,\dots,p_m\})\,|{\cal M}_n(\{P,p_{m+1},\dots, p_n\})\rangle + \dots \,.
\end{equation}
Since (\ref{30}) is valid for both the dimensionally regularized amplitude as well as the minimally subtracted amplitude, it is possible to derive a renormalization-group equation for the splitting amplitude ${\bf Sp}(\{p_1,\dots,p_m\})$. Applying charge conservation, and noting that the splitting amplitude commutes with partons not involved in the splitting process, one obtains
\begin{equation}\label{33}
   \frac{d}{d\ln\mu}\,{\bf Sp}(\{p_1,\dots,p_m\},\mu) 
   = {\bf\Gamma}_{\rm Sp}(\{p_1,\dots,p_m\},\mu)\,{\bf Sp}(\{p_1,\dots,p_m\},\mu) \,,
\end{equation}
where we have introduced
\begin{equation}\label{34}
   {\bf\Gamma}_{\rm Sp}(\{p_1,\dots,p_m\},\mu) 
   = {\bf\Gamma}(\{p_1,\dots,p_n\},\mu)
    - {\bf\Gamma}(\{P,p_{m+1},\dots,p_n\},\mu) \Big|_{\T_P=\T_1+\dots+\T_m} \,,
\end{equation}
with $\T_P$ the color generator associated with the parent parton $P$. In analogy with the two-particle collinear limit, the anomalous dimension of the splitting function must be independent of the color generators and momenta of the partons not involved in the splitting process. 

It is interesting to investigate whether the multi-particle collinear limit gives additional constraints on the functions $F$ in (\ref{eq8a}) and $G_{1,2,3,4}$ in (\ref{dG4fin}). Unfortunately, we finds that no additional constraints are obtained on those functions ($F$ and $G_{1,2,3}$) that involve only four external partons. This can be seen considering as an example the function $F$. With up to four partons involved it makes sense to consider the three-particle collinear limit only. We parameterize the three momenta which become collinear as 
\begin{equation}
   p_i^\mu = z_i E n^\mu + p_{\perp,i}^\mu - \frac{p_{\perp,i}^2}{4z_iE}\,\bar n^\mu \,; 
   \qquad i=1,2,3 \,, 
\end{equation}
with $z_1+z_2+z_3=1$ and $p_{\perp,1}+p_{\perp,2}+p_{\perp,3}=0$. Here $n$ and $\bar n$ are two light-like vectors, satisfying $n\cdot\bar n=2$. The momentum fractions $z_i$ as well as the transverse momenta $p_{\perp,i}$ are assumed to be of the same order. In this way $p_{\perp,i}\sim p_\perp$, and $p_\perp/E$ is a small expansion parameter. The collinear limit corresponds to taking $p_\perp\to 0$ at fixed $E$. With these definitions, to first order in $p_\perp/E$ we have $-s_{ij}=-(p_i+p_j)^2=(z_i p_{\perp,j}-z_j p_{\perp,i})^2/(z_iz_j)\equiv z_i z_j\lambda_{ij}$. We then find
\begin{equation}
\begin{aligned}
   \ep_{ij} \equiv \beta_{1ij2} 
   &= \frac{1}{z_1 E} \left( \frac{p_{\perp,1}\cdot p_i}{n\cdot p_i}
    - \frac{p_{\perp,1}\cdot p_j}{n\cdot p_j} \right) 
    - \frac{1}{z_2 E} \left( \frac{p_{\perp,2}\cdot p_i}{n\cdot p_i} 
    - \frac{p_{\perp,2}\cdot p_j}{n\cdot p_j} \right) \to 0 \,, \\
   \omega_{ij} \equiv \beta_{12ij} 
   &= \ln\frac{\lambda_{12}}{4 E^2} + \ln\frac{(-s_{ij})}{(-n\cdot p_i)(-n\cdot p_i)} 
    \to -\infty \,,
\end{aligned}
\end{equation}
and similar results hold for $\beta_{1ij3}$, $\beta_{13ij}$, $\beta_{2ij3}$, and $\beta_{23ij}$. In addition, we obtain
\begin{equation}
   \beta_{123i} = \ln\frac{(-s_{12})(-s_{3i})}{(-s_{13})(-s_{2i})}
    = \ln \frac{\lambda_{12}}{\lambda_{13}} \,, \qquad
   \beta_{13i2} - \beta_{1i23} = \ln\frac{\lambda_{12}\lambda_{13}}{\lambda_{23}^2} \,.
\end{equation}
According to the definition (\ref{34}), the contribution of the four-particle correlation term proportional to $F$ in (\ref{eq8a}) to the anomalous dimension of the splitting amplitude reads
\begin{equation}\label{37}
\begin{aligned}
   {\bf\Delta \Gamma}_{\rm Sp}(\{p_1,p_2,p_3,\mu\}) 
   &= \bigg\{ \sum_{i,j\neq 1,2,3} \Big[ 8 {\cal T}_{12ij}\,F(\beta_{12ij}, \beta_{12ij}) 
    + 4 {\cal T}_{1ij2}\,F(\beta_{1ij2},-2\beta_{12ij}) \Big] \\
   &\hspace{1.0cm}\mbox{}+ (12\leftrightarrow 13) + (12\leftrightarrow 23) \bigg\} \\
   &\hspace{-2.0cm}\mbox{}+ 8\!\sum_{i\neq 1,2,3} \bigg[ {\cal T}_{123i}\,
    F\bigg(\ln\frac{\lambda_{12}}{\lambda_{13}},
           \ln\frac{\lambda_{12}\lambda_{13}}{\lambda_{23}^2}\bigg)
    + {\cal T}_{213i}\,F\bigg(\ln\frac{\lambda_{12}}{\lambda_{23}},
                              \ln\frac{\lambda_{12}\lambda_{23}}{\lambda_{13}^2}\bigg) \\ 
   &\hspace{-1mm}\mbox{}+ {\cal T}_{312i}\,F\bigg(\ln\frac{\lambda_{13}}{\lambda_{23}},
                                     \ln\frac{\lambda_{13}\lambda_{23}}{\lambda_{12}^2}\bigg)
    \bigg] \,.
\end{aligned}
\end{equation}
The terms in the first two lines are a generalization of the term appearing in the two-particle collinear limit, as can be seen by comparing with relations~(101) and (102) in \cite{Becher:2009qa}. Since it involves external partons other than 1, 2, 3, it must vanish in all collinear limits. The conditions for this to happen are precisely the same as in the case of two collinear particles. The other terms in (\ref{37}) instead do not depend on kinematic variables involving momenta other than those of particles 1, 2, 3. One can therefore use the color-conservation relation (\ref{colorcons}) to perform the sum over the free index $i$ and express all color structures in terms of structures involving the indices of the three collinear particles only. We conclude that the terms in the last two lines are compatible with the three-particle collinear limit, and no new constraints are obtained.

The functional form of the five-particle correlation $G_4$, on the other hand, {\em can\/} be constrained by both the two- and three-particle collinear limits. For instance, in the case of the two-particle collinear limit one finds that the independent color structures ${\cal T}_{12ijk}$, ${\cal T}_{i12jk}$, and ${\cal T}_{1ijk2}$ contribute to ${\bf\Delta\Gamma}_{\rm Sp}(\{p_1,p_2\},\mu)|_{G_4}$. The coefficients multiplying these three structures must vanish in the collinear limit. For example, one obtains
\begin{equation}\label{134}
\begin{aligned}
   {\cal T}_{12ijk} \Big[
   & G_4(-\omega_{ij},\omega_{ij},\beta_{i1jk},\beta_{ijk1},-\omega_{ik})
    - G_4(-\omega_{ij},\omega_{ij},-\omega_{ik},\omega_{ik},\beta_{1ikj}) \\
   &\mbox{}- G_4(\omega_{ij},\epsilon_{ij},\beta_{ij1k},\beta_{i1kj},\beta_{ijk2})
    + G_4(\omega_{ij},\epsilon_{ij},\omega_{ik},\epsilon_{ik},\omega_{jk}) \Big]
    + \dots = 0 \,.
\end{aligned}
\end{equation}
In the case of the three-particle collinear limit, one has five additional independent color structures contributing to ${\bf\Delta\Gamma}_{\rm Sp}(\{p_1,p_2,p_3\},\mu)|_{G_4}$, namely ${\cal T}_{123ij}$ ${\cal T}_{213ij}$, ${\cal T}_{312ij}$, ${\cal T}_{12ij3}$, and ${\cal T}_{1ij23}$, whose coefficients must also vanish. For instance
\begin{equation}\label{135}
   {\cal T}_{123ij}\,G_4(\beta_{123i},\beta_{13i2},\beta_{123j},\beta_{13j2},\beta_{12ji})
   + \dots = 0 \,.
\end{equation}
The form of the remaining constraints is similar to (\ref{134}) and (\ref{135}). Since they are
not particularly illuminating, we do not report them here.


\begin{thebibliography}{99}

\bibitem{Becher:2009cu} 
  T.~Becher and M.~Neubert,
  Phys.\ Rev.\ Lett.\  {\bf 102}, 162001 (2009)
  [arXiv:0901.0722 [hep-ph]].

\bibitem{Korchemskaya:1994qp} 
  I.~A.~Korchemskaya and G.~P.~Korchemsky,
  Nucl.\ Phys.\ B {\bf 437}, 127 (1995)
  [hep-ph/9409446].

\bibitem{Catani:1996vz} 
  S.~Catani and M.~H.~Seymour,
  Nucl.\ Phys.\ B {\bf 485}, 291 (1997)
  [Erratum-ibid.\ B {\bf 510}, 503 (1998)]
  [hep-ph/9605323].

\bibitem{Becher:2003kh} 
  T.~Becher, R.~J.~Hill, B.~O.~Lange and M.~Neubert,
  Phys.\ Rev.\ D {\bf 69}, 034013 (2004)
  [hep-ph/0309227].

\bibitem{Korchemskaya:1992je} 
  I.~A.~Korchemskaya and G.~P.~Korchemsky,
  Phys.\ Lett.\ B {\bf 287}, 169 (1992).

\bibitem{Bauer:2001yt} 
  C.~W.~Bauer, D.~Pirjol and I.~W.~Stewart,
  Phys.\ Rev.\ D {\bf 65}, 054022 (2002)
  [hep-ph/0109045].
  
\bibitem{Gardi:2009qi} 
  E.~Gardi and L.~Magnea,
  JHEP {\bf 0903}, 079 (2009)
  [arXiv:0901.1091 [hep-ph]].

\bibitem{Becher:2009qa} 
  T.~Becher and M.~Neubert,
  JHEP {\bf 0906}, 081 (2009)
  [arXiv:0903.1126 [hep-ph]].

\bibitem{Gatheral:1983cz} 
  J.~G.~M.~Gatheral,
  Phys.\ Lett.\ B {\bf 133}, 90 (1983).

\bibitem{Frenkel:1984pz} 
  J.~Frenkel and J.~C.~Taylor,
  Nucl.\ Phys.\ B {\bf 246}, 231 (1984).

\bibitem{Gardi:2010rn} 
  E.~Gardi, E.~Laenen, G.~Stavenga and C.~D.~White,
  JHEP {\bf 1011}, 155 (2010)
  [arXiv:1008.0098 [hep-ph]].

\bibitem{Berends:1988zn} 
  F.~A.~Berends and W.~T.~Giele,
  Nucl.\ Phys.\ B {\bf 313}, 595 (1989).

\bibitem{Mangano:1990by} 
  M.~L.~Mangano and S.~J.~Parke,
  Phys.\ Rept.\  {\bf 200}, 301 (1991)
  [hep-th/0509223].

\bibitem{Bern:1995ix} 
  Z.~Bern and G.~Chalmers,
  Nucl.\ Phys.\ B {\bf 447}, 465 (1995)
  [hep-ph/9503236].

\bibitem{Kosower:1999xi} 
  D.~A.~Kosower,
  Nucl.\ Phys.\ B {\bf 552}, 319 (1999)
  [hep-ph/9901201].

\bibitem{Armoni:2006ux} 
  A.~Armoni,
  JHEP {\bf 0611}, 009 (2006)
  [hep-th/0608026].

\bibitem{Alday:2007hr} 
  L.~F.~Alday and J.~M.~Maldacena,
  JHEP {\bf 0706}, 064 (2007)
  [arXiv:0705.0303 [hep-th]].

\bibitem{Alday:2007mf} 
  L.~F.~Alday and J.~M.~Maldacena,
  JHEP {\bf 0711}, 019 (2007)
  [arXiv:0708.0672 [hep-th]].

\bibitem{Dixon:2009ur} 
  L.~J.~Dixon, E.~Gardi and L.~Magnea,
  JHEP {\bf 1002}, 081 (2010)
  [arXiv:0910.3653 [hep-ph]].

\bibitem{Bret:2011xm} 
  V.~Del Duca, C.~Duhr, E.~Gardi, L.~Magnea and C.~D.~White,
  Phys.\ Rev.\ D {\bf 85}, 071104 (2012)
  [arXiv:1108.5947 [hep-ph]].
  
\bibitem{DelDuca:2011ae} 
  V.~Del Duca, C.~Duhr, E.~Gardi, L.~Magnea and C.~D.~White,
  JHEP {\bf 1112}, 021 (2011)
  [arXiv:1109.3581 [hep-ph]].

\bibitem{Balitsky:1979ap} 
  I.~I.~Balitsky, L.~N.~Lipatov and V.~S.~Fadin,
  in {\em Leningrad 1979, Proceedings, Physics Of Elementary Particles}, Leningrad 1979, 109-149 
  (in Russian)

\bibitem{Bogdan:2006af} 
  A.~V.~Bogdan and V.~S.~Fadin,
  Nucl.\ Phys.\ B {\bf 740}, 36 (2006)
  [hep-ph/0601117].
  
\bibitem{Fadin:2006bj} 
  V.~S.~Fadin, R.~Fiore, M.~G.~Kozlov and A.~V.~Reznichenko,
  Phys.\ Lett.\ B {\bf 639}, 74 (2006)
  [hep-ph/0602006].

\bibitem{Korchemskaya:1996je} 
  I.~A.~Korchemskaya and G.~P.~Korchemsky,
  Phys.\ Lett.\ B {\bf 387}, 346 (1996)
  [hep-ph/9607229].

\bibitem{Dokshitzer:2005ig} 
  Y.~.L.~Dokshitzer and G.~Marchesini,
  JHEP {\bf 0601}, 007 (2006)
  [hep-ph/0509078].

\bibitem{Bern:2008pv}
  Z.~Bern, J.~J.~M.~Carrasco, L.~J.~Dixon, H.~Johansson and R.~Roiban,
  Phys.\ Rev.\  D {\bf 78}, 105019 (2008)
  [arXiv:0808.4112 [hep-th]].

\bibitem{Bern:2006ew} 
  Z.~Bern, M.~Czakon, L.~J.~Dixon, D.~A.~Kosower and V.~A.~Smirnov,
  Phys.\ Rev.\ D {\bf 75}, 085010 (2007)
  [hep-th/0610248].

\bibitem{Catani:2011st} 
  S.~Catani, D.~de Florian and G.~Rodrigo,
  JHEP {\bf 1207}, 026 (2012)
  [arXiv:1112.4405 [hep-ph]].

\bibitem{Forshaw:2012bi} 
  J.~R.~Forshaw, M.~H.~Seymour and A.~Siodmok,
  arXiv:1206.6363 [hep-ph].

\bibitem{Moch:2004pa} 
  S.~Moch, J.~A.~M.~Vermaseren and A.~Vogt,
  Nucl.\ Phys.\ B {\bf 688}, 101 (2004)
  [hep-ph/0403192].

\bibitem{Catani:2003vu} 
  S.~Catani, D.~de Florian and G.~Rodrigo,
  Phys.\ Lett.\ B {\bf 586}, 323 (2004)
  [hep-ph/0312067].

\end{thebibliography}
\end{document}